\documentclass[reprint,superscriptaddress,amsmath,amssymb,aps]{revtex4-1}
\usepackage[utf8]{inputenc}
\usepackage{graphicx}
\usepackage{dcolumn}
\usepackage{bm}
\usepackage[colorlinks,citecolor=blue,linkcolor=blue,urlcolor=blue]{hyperref}
\usepackage{ulem}

\begin{document}

\title{Spectral functions of the honeycomb lattice with both the Hubbard and long-range Coulomb Interactions}

\author{Ho-Kin Tang}
\affiliation{School of Science, Harbin Institute of Technology, Shenzhen, 518055, China}
\affiliation{ Centre for Advanced 2D Materials, National
   University of Singapore, 6 Science Drive 2, Singapore 117546.}
\author{Indra Yudhistira}
 \affiliation{ Centre for Advanced 2D Materials, National
   University of Singapore, 6 Science Drive 2, Singapore 117546.}
 \affiliation{ Department of Physics, Faculty of Science, National University
   of Singapore, 2 Science Drive 3, Singapore 117542.}
   \author{Udvas Chattopadhyay} 
   \affiliation{ Centre for Advanced 2D Materials, National
   University of Singapore, 6 Science Drive 2, Singapore 117546.}
   \affiliation{ Yale-NUS College, 16 College Ave West, Singapore 138527.}
 \author{Maksim Ulybyshev}
 \affiliation{ Institut f\"ur Theoretische Physik und Astrophysik, Universit\"at W\"urzburg, 
   Am Hubland, D-97074 W\"urzburg, Germany.}
 \author{P. Sengupta}
 \affiliation{ Centre for Advanced 2D Materials, National
   University of Singapore, 6 Science Drive 2, Singapore 117546.}
 \affiliation{ School of Physical and Mathematical Sciences, Nanyang Technological 
   University, 21 Nanyang Link, Singapore 637371.}
 \author{F. F. Assaad}
 \affiliation{ Institut f\"ur Theoretische Physik und Astrophysik, Universit\"at W\"urzburg, 
   Am Hubland, D-97074 W\"urzburg, Germany.}
 \author{S. Adam}
 \affiliation{ Centre for Advanced 2D Materials, National
   University of Singapore, 6 Science Drive 2, Singapore 117546.}
 \affiliation{ Department of Physics, Faculty of Science, National University
   of Singapore, 2 Science Drive 3, Singapore 117542.}
 \affiliation{ Yale-NUS College, 16 College Ave West, Singapore 138527.}
 \affiliation{Department of Materials Science and Engineering, 
 National University of Singapore, 9 Engineering Drive 1, 
 Singapore 117575}
\date{\today}

\begin{abstract}
The   absence  of screening of the non-local Coulomb interaction in Dirac systems at  charge  neutrality  leads  to the  breakdown  of  the Fermi liquid and   divergence of  the  Fermi velocity.   On the other  hand, Mott  Hubbard physics   and the concomitant  formation of local  moments  is  dominated  by  the   local  effective  Hubbard interaction.  Using quantum Monte  Carlo  methods  combined  with  stochastic analytical continuation,  we compute  the single particle  spectral  function  of fermions on the   honeycomb lattice  for a realistic interaction that includes both the  Hubbard  interaction  and  long-ranged  Coulomb  repulsion.  To  a  first approximation, we  find  that  the  generic  high-energy  features  such as the formation of the upper  Hubbard band  are   independent  of the  long-ranged  Coulomb  repulsion and determined mostly by the local  effective    Hubbard interaction.  The sub-leading effects  of adding the long-range interaction include an  enhancement of  the bandwidth  and a decrease of the spin-polaron quasi-particle weight and lifetime.     
\end{abstract}
\pacs{Valid PACS appear here}

\maketitle
\textit{Introduction -- } Beyond being of fundamental intrinsic interest, the Hubbard model is held by many to contain the essential mechanisms that explain unconventional and high-temperature superconductivity~\cite{Hubbard1963-ot,Gutzwiller1964-cb,Tasaki1998-le, Norman2011-xi,Sigrist1991-uu,Bednorz1986-ro,Proust2019-pv,Anderson2007-ss,Sachdev2003-ks, Dagotto1994-qt,Norman2003-bz,Isobe2018-in}. Despite years of study, there is still no consensus on the solution to the Hubbard model away from charge neutrality.  At weak coupling, the Hubbard model hosts a Landau Fermi liquid with a quasiparticle band set by the lattice geometry. In the strong coupling limit, the density of states splits into two bands called the upper and the lower Hubbard bands and the occurrence of these bands is many-body in nature. Between these two limits, the competition between these two physical pictures leads to many novel phenomena, like the Mott-insulator transition and charge or spin density waves (see e.g. Ref. \onlinecite{Sachdev2003-ks}). The appearance of the upper Hubbard band indicates the onset of strong coupling many-body phenomena in the Hubbard model.

In recent years there have been attempts to simulate the Hubbard model in trapped atomic systems. Despite being at nano Kelvin temperatures these systems are still unable to probe the low-temperature properties of the model~\cite{bloch_quantum_2012}.  With the recent observation of superconductivity in twisted bilayer graphene and related systems~\cite{Cao2018-Unconventional, Lu2019-Superconductors,Saito2020-Independent,Stepanov2021-Competing} and explicit mappings of twisted transition metal dichalcogenides as Hubbard model simulators~\cite{xu_tunable_2022, wu_hubbard_2018}, lattice geometries beyond the square one are now experimentally relevant.  Unlike the cold atom lattices where the interactions are largely on-site, in these two-dimensional materials, the long-range non-local component of the Coulomb interaction is essential~\cite{Das_Sarma2011-cq,Kotov2012-rb,Schuler2013-ys}.  In particular, the non-local part of the Coulomb interaction effectively alters the transport properties and spectroscopic properties seen in magnetotransport~\cite{Elias2011-ry}, infrared spectroscopy~\cite{Li2008-ej}, and the Raman scattering~\cite{Faugeras2015-wi} experiments.  The long-range nature of the Coulomb interaction also alters the many-body properties such as shifting instability of the charge and spin in phase diagram~\cite{Tang2018-au, Ulybyshev2013-sw,Protsenko2021-re}.  In light of new experiments in 2D materials, it is important to understand the properties of the upper Hubbard band in presence of a non-local Coulomb interaction.

The upper Hubbard band has been studied extensively for the Hubbard model on a square lattice~\cite{Preuss1995-aj,Preuss1997-hz,Brunner2000-rh,Kyung2006-nf,Macridin2007-gg,Zemljic2008-fw,Sakai2010-ra,Rost2012-sc,Kohno2014-vc,Yang2016-sr,Wang2018-rg}.  These results seem relevant to strongly correlated phenomena in cuprates where both theory and experiment observe the upper Hubbard band and the `waterfall' phenomena~\cite{Preuss1995-aj,Moritz2010-mc,Kampf1990-oh,Graf2007-jf}.  While the phase transitions and quasiparticle properties of the Hubbard model on a honeycomb lattice have now also been established (see e.g. Refs.~\cite{Geim2009-bf,Castro_Neto2009-ee,Meng2010-lf,Sorella1992-rw,Sorella2012-cy,Assaad2013-pr,Wu2014-jg,Herbut2009-ip,Herbut2006-jf}), to our knowledge, the properties of the upper Hubbard band in the presence of a long-range Coulomb interaction has not previously been discussed.  In this work we use the unbiased quantum Monte Carlo method~\cite{assaad2008world} since it is known that other methods to study strongly correlated systems such as dynamical mean-field theory are unable to resolve spin  fluctuations  and  associated  features  such  as so  called  water-falls  in the spectral function~\cite{Raczkowski2020-fg,Raczkowski2021-uq}. Although we are restricted to the half-filled case because of the sign problem~\cite{Loh1990-xm}, this situation can readily be probed in two-dimensional quantum simulators. More generally, we hope that solution at half-filing can provide some insight into the doped case.  

\begin{figure*}[ht!]
    \centering
    \includegraphics[width=0.9\linewidth]{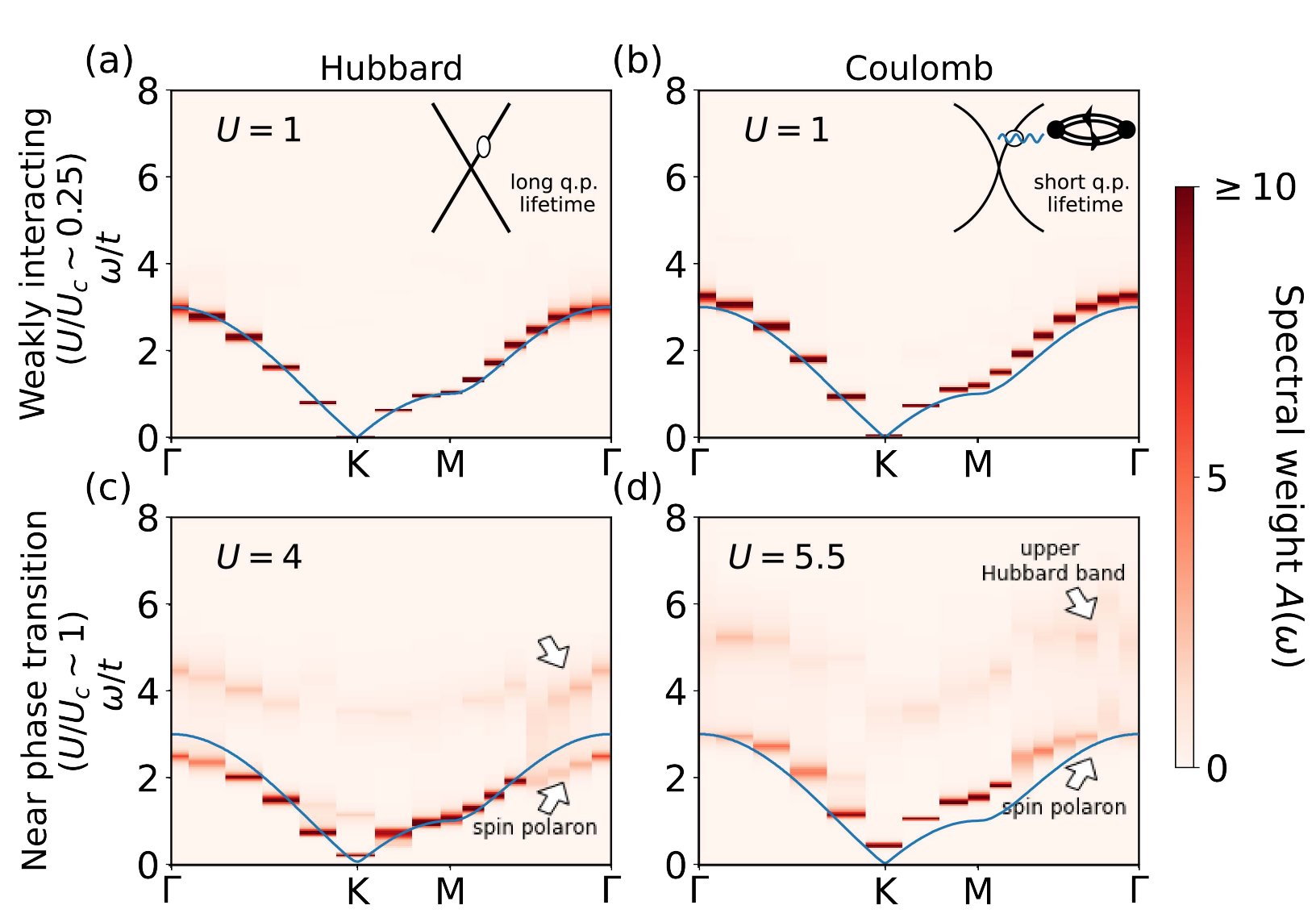}
    \caption{The spectral function for fermions on a honeycomb lattice with and without a long-range Coulomb interaction.  Left panels show a pure Hubbard model, while right panels include the non-local interaction.  Top panels (a) and (b) are the weakly interacting regime; and bottom panels (c) and (d) are close to the phase transition.  We observe that close to the phase transition, two distinguishable bands of spectral weight appear near the $\Gamma$ point, which we identify as the coherent spin polaron band and the upper Hubbard band. We also observe that the Coulomb interaction increases the bandwidth of the quasiparticle band, diminishes the lifetime and transfers more weight to the incoherent part of the spectral.  Solid lines show the mean-field theory results~\protect{\cite{supp}}. For   the  Hubbard  model  $U_c/t  \simeq  3.8$ while for the long-range Coulomb interaction, we have $U_c/t \simeq  5.5$.}. 
    \label{fig:fig1}
\end{figure*}

\textit{Model and Methods -- }  We study the spin-half Hubbard model with the Coulomb interaction on a honeycomb lattice. The Hamiltonian is given as
\begin{eqnarray*}
\hat{H} = - t \sum_{\langle i j \rangle, \sigma} \big( \hat{c}_{i\sigma}^{\dagger} \hat{c}_{j\sigma} + h.c. \big )
+ \frac{1}{2} \sum_{i, j} (\hat{n}_{i}-1) \mathcal{V}_{i j} (\hat{n}_{j}-1) \,
\label{eq:ExtHubbmodelH}
\end{eqnarray*}
where $\hat{c}_{i\sigma}^{\dagger}$ ($\hat{c}_{i\sigma}$) creates (annihilates) an electron of spin $\sigma=\uparrow\downarrow$ at the atom located at position $\mathbf{r}_{i}$. The operator $\hat{n}_i=\sum_{\sigma}\hat{n}_{i\sigma}=\sum_{\sigma}\hat{c}_{i\sigma}^{\dagger}\hat{c}_{i\sigma}$ counts the number of electrons with spin $\sigma$ sitting at the atom located at position $\mathbf{r}_{i}$. The nearest-neighbor hopping integral is $t$ that defines our energy scale, while $\mathcal{V}_{ij}$ stands for the Coulomb interaction potential between two electrons sitting at sites $\mathbf{r}_{i}$ and $\mathbf{r}_{j}$. For the Hubbard model case, we set $\mathcal{V}_{ij}=U t \delta_{ij}$. With the non-local Coulomb interaction, we set the diagonal part as $Ut$ and the off-diagonal part as $r_{\delta} U t/2r_{ij}$, where $r_\delta$ is the nearest neighbor distance in the honeycomb lattice. 

The phase diagram of the honeycomb Hubbard model has been well studied.  The increase of Hubbard interaction leads to the phase transition from the semi-metal phase to the anti-ferromagnetic phase.  This phase transition is in the Gross-Neveu universality class.  The inclusion of the non-local Coulomb interaction increases $U_c$, the critical value of Hubbard U required for this transition, but does not change the universality class~\cite{Tang2018-au}.  

To compare the local and non-local models, we pick values of Hubbard U with comparable values of $U/U_c$.  In this work, we compare the spectral functions both with and without the non-local interactions.  To do this, we first obtain the retarded Green's function of the ground state along the imaginary time by the quantum Monte Carlo~\cite{assaad2008world,Tang2018-au,ALF_v2}, then use the stochastic maximum entropy method~\cite{Sandvik1998-nz} to process the Green's function and obtain the spectral function $A(\omega ,k)$ of the system with the frequency $\omega$ and the momentum $k$. 

In the Supplementary Material~\cite{supp} we provide details of how we do the stochastic maximum entropy processing and Lorentzian fitting of $A(\omega ,\Gamma )$ to obtain the excitation energy, spectral weight, and quasiparticle lifetime.  For the rest of the paper we discuss our findings.  

\begin{figure}[tb!]
    \centering
    \includegraphics[width=1\linewidth]{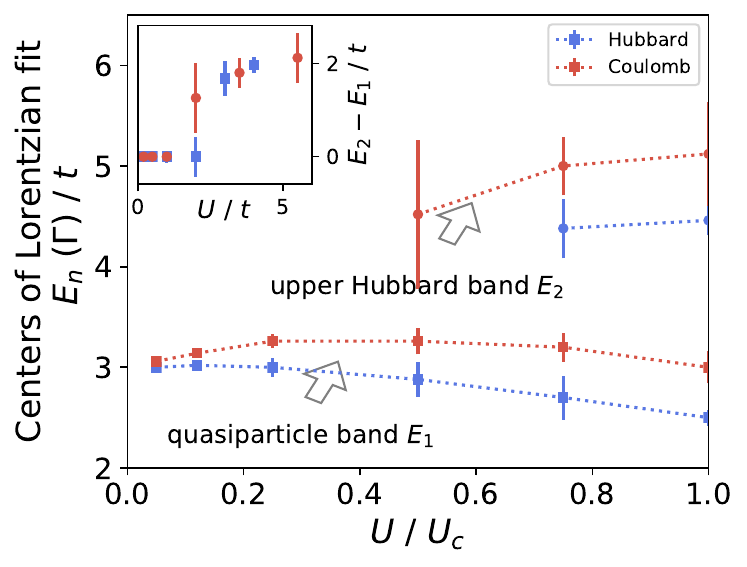}
    \caption{The frequency of the quasiparticle band and the upper Hubbard band at the $\Gamma$ point, estimated by the Lorentzian fitting method discussed in the supplementary material~\cite{supp}. The upper Hubbard band appears in the middle of the semi-metal phase. In the inset, we find that the energy difference between two bands mostly depend on Hubbard $U$.}
    \label{fig:fig2}
\end{figure}

\textit{Results and Discussion -- }  Our main results are shown in Fig.~\ref{fig:fig1} where we compare the pure Hubbard model (left panels) with the realistic Coulomb potential including the long-range tail (right panels).  The top panels have relatively weak interactions. For  the Hubbard   we  find  that a  paramagnetic   mean-field solution (see Supplemental material for details) adequately captures the effects of interactions.      In the presence of the long-ranged  Coulomb interaction  deviations  for  the paramagnetic  mean-field  result  are  expected.   First,  we  expect  the  velocity  to  diverge  \cite{Tang2018-au,Ulybyshev21}  logarithmically   as  one  approaches  the  Fermi  energy.  This  low  energy  effect   lies  beyond  our  energy   resolution. Second,  we  observe   a  slight   enhancement of  the  band-width.   This high  energy  effect can  be captured   at  the  mean-field  level by   writing  the   nearest  neighbor  repulsion as:   
$   V  \left(\hat{n}_i -1  \right)   \left( \hat{n}_j -1 \right) = - V \left(  \sum_{\sigma} \hat{c}_{i\sigma}^{\dagger} \hat{c}_{j\sigma} + h.c. \right )^2 -  4V 
   \hat{\mathbf{S}}_i  \cdot \hat{\mathbf{S}}_j  +  4 V \left(  \Delta^{\dagger}_{i}\Delta^{\phantom\dagger}_{j}  +    h.c. \right) $.  Here,  the  equality  holds  up to  a constant, where $\Delta^{\dagger}_{i} =  \hat{c}^{\dagger}_{i,\uparrow} \hat{c}^{\dagger}_{i,\downarrow}  $   and   $ \hat{\mathbf{S}}_i $ is  the spin  operator.   A  mean  field  approach   for  a  paramagnetic  state  can be   characterized  by:  $ \langle \Delta^{\dagger}_{i}\rangle = 
 \langle  \hat{\mathbf{S}}_i     \rangle  = 0 $,  but   $\langle \sum_{\sigma} \hat{c}_{i\sigma}^{\dagger} \hat{c}_{j\sigma} + h.c. \rangle  =  \chi $.  Thereby   the  nearest   neighbor   repulsion  interaction  can  account  for  the   observed   band   renormalization  and   weak  coupling  (see Fig.~\ref{fig:fig2}).  We have also confirmed this interpretation using lattice Random Phase Approximation calculations for the renormalization of the energy levels by comparing two cases: one with the full Coulomb tail, and another one with only on-site and nearest-neighbour interactions. The shift of the energy levels near the Dirac point decreases in the latter case, which is expected since the long-range Coulomb tail is mainly responsible for the logarithmic divergence of the Fermi velocity. However, the shift of the energy levels at $\Gamma$-point increases in the computation with only Hubbard and nearest-neighbour repulsion. Thus we attribute the observed broadening of the bandwidth mainly to the nearest-neighbour interaction, with the longer range Coulomb tail working against this effect.

We now discuss the pure Hubbard model results (two left panels of Fig.~\ref{fig:fig1}).  We note that the spectral features close to the quantum phase transition are significantly different from those in the weakly interacting regime.   Most notably, the single quasi-particle band splits into two distinguishable bands as we approach the phase transition.  For the pure Hubbard model our results are consistent with the previous studies on both the square and honeycomb lattices~\cite{Preuss1995-aj,Preuss1997-hz,Brunner2000-rh,Kyung2006-nf,Macridin2007-gg,Zemljic2008-fw,Sakai2010-ra,Rost2012-sc,Kohno2014-vc,Yang2016-sr,Wang2018-rg,Sorella1992-rw,Sorella2012-cy,Assaad2013-pr,Wu2014-jg,Herbut2009-ip,Herbut2006-jf}.  For completeness, we first summarize our findings for the pure Hubbard model before considering the addition of the long-range Coulomb interaction.

\begin{figure}[tb!]
    \centering
    \includegraphics[width=1\linewidth]{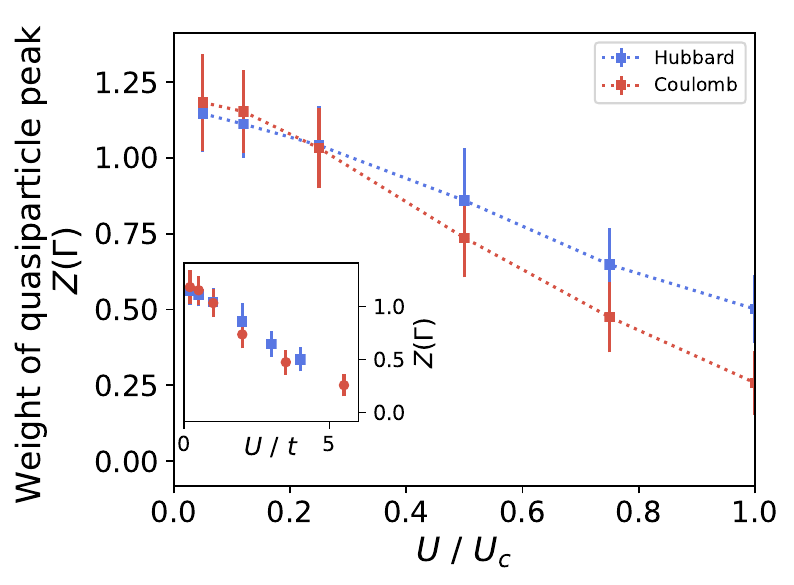}
    \caption{The upper bound of the weight $Z$ of the quasiparticle band at $\Gamma$, estimated by integrating the Lorentzian    used  to  fit the  data.  
    The  sum  rule  is set  by   $\int_{0}^{\infty} d \omega A(\omega) = 1 $ such  that  $Z$  should   approach  this  value at  small couplings.  The  deviation 
    is  a  consequence  of  the  fit  procedure.   In the inset, we plot $Z$ against the Hubbard $U$. The weight shows almost a linear drop with $U$. On this ground, the long-range Coulomb interaction further diminishes $Z(\Gamma$). }
    \label{fig:fig3}
\end{figure}

The higher energy feature  that  we  will   refer to   as  the  upper  Hubbard band  is only visible at sufficiently strong Hubbard $U$ and its spectral weight increases with increasing $U$. Moreover, its weight spreads within a frequency range indicating that the upper Hubbard band is not a coherent quasiparticle.  When the interactions are strong enough, the quasiparticle band then dissolves into an incoherent background. In Fig.~\ref{fig:fig2} we plot the centers of the Lorentzian fits for both the upper Hubbard band (E$_2$) and the quasiparticle band (E$_1$). In the inset to Fig.~\ref{fig:fig2}, we show that the energy difference between quasiparticle peak and the center of the upper Hubbard band collapses as function of $U$.  This shows that the upper Hubbard band is highly correlated with the magnitude of the Hubbard $U$, and in the strong coupling limit, it survives even as the quasiparticle weight diminishes as shown in Fig.~\ref{fig:fig3}.  We also note that these features of the upper Hubbard band are in agreement with a recent findings from the instanton gas approximation \cite{PhysRevB.107.045143}; in particular the appearance of the upper Hubbard band at around $U/U_c=0.9$, and the gap between upper and lower bands increasing with $U$.  This reiterates that the appearance of the upper band can serve as a manifestation of the importance of non-mean field saddle points in the configuration space of the Hubbard model in the strong coupling limit.

Consistent with what is known in the literature for the square-lattice, we label the lower band (E$_1$) as the coherent spin-polaron. The spin-polaron band represents an microscopic picture of a hole propagating in the background of antiferromagnetic fluctuations where the bandwidth is proportional to the magnetic scale i.e. $J\approx t^2/U$.  Indeed, we find the decreasing trend of $E_1(\Gamma)$ in the pure Hubbard case where the decrease is around around 17$\%$ at the phase transition.

We now consider the results including the non-local Coulomb interaction (right two panels in Fig.~\ref{fig:fig1}).  The main conclusion is that the evolution of the spectrum with non-local Coulomb interaction shows similar qualitative behavior to that of the pure Hubbard model.  This is  consistent  with the notion that the physics close to the anti-ferromagnetic phase transition remains controlled by local interactions.   In what follows, we look more closely to identify the differences arising from the long-range interactions.  We choose to focus on the properties of the spectral function at the $\Gamma$ point.  This is because we find that the $\Gamma$ point properties are relatively insensitive to lattice sizes, in contrast to the K-point, where properties like the spectral gap are strongly influenced by finite-size effects~\cite{Tang2019-ty}.  

 Looking more closely at the data, three differences are observable: (i) The non-local Coulomb interaction shifts both peaks to higher frequency. 
 This is shown more clearly in Fig.~\ref{fig:fig2} where Lorentzian centers of the spectral functions with the non-local interaction (red points) are systematically larger than those with only a pure Hubbard interactions (blue points). This is  consistent  with the    aforementioned    enhancement  of  the   bandwidth.   
 (ii) With the long-range Coulomb interaction, there is more spectral weight transferred to the upper Hubbard band.  As a consequence, the upper Hubbard band is visible at lower effective values of Hubbard interaction ($U/U_c \sim 0.5$) compared to the pure Hubbard model.   
  (iii) There is a widening of the peaks in both the upper Hubbard band and the spin polaron bands.  To further quantify this we show in Fig.~\ref{fig:fig3} the weight in the quasiparticle peak $Z(\Gamma)$ as a function of Hubbard $U$.  Consistent with the other observables discussed above, in the inset we show that the dominant factor for the decrease of the quasiparticle weight is  the local Hubbard $U$. The effect of the long-range Coulomb interaction is to further diminish the weight (effectively by shifting the critical Hubbard $U$ to a larger value).  This gives an intuitive way to understand our findings:  For a fixed $U/U_c$, the larger value of $U_c$ arising from the inclusion of the long-range Coulomb tail, also implies a larger $U$ -- the effect of which can qualitatively explain the larger spectral weight, higher bandwidth and diminished quasiparticle weight for the realistic Coulomb interaction relative to the pure Hubbard model.

\begin{figure}[tb!]
    \centering
    \includegraphics[width=1\linewidth]{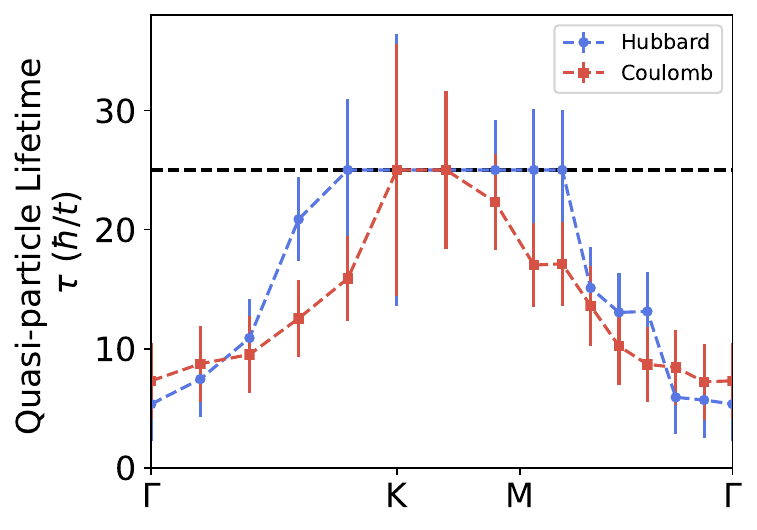}
    \caption{The qualitative estimation of the quasiparticle lifetime $\tau$ along the high symmetry point of graphene. The weakly interacting result~($U=1$) is plotted (blue) with pure Hubbard interaction and (red) with Coulomb interaction. The long-range Coulomb interaction shortens the quasiparticle lifetime. The lifetime is estimated by $\tau=1/(2\Gamma(k))$, where $\Gamma$ is the width of the spectral peak. The dashed line is the upper bound of the estimation set by our spectral interval in frequency~(25 $\hbar/t$). }
    \label{fig:fig4}
\end{figure}

Finally, we look at the quasiparticle lifetime which provides additional information e.g. on the electronic transport properties of the material. Our simulation only considers the electron-electron scattering by the local Hubbard and non-local Coulomb interaction.  The lifetime can be estimated by the inverse of the width of the quasiparticle peak.  Unfortunately, our lifetime calculation is limited by the interval of the spectral function (with maximum bound of about 25 $\hbar/t$ in our case).  Nonetheless, its qualitative behavior is worth investigating. We show our result in Fig.~\ref{fig:fig4}. The lifetime decreases with the increasing distance from the Fermi energy. Compared to the pure Hubbard model, the long-range Coulomb interaction induces drops in the lifetime in most regions of momentum space.

\textit{Conclusion -- }  In this work we used an unbiased quantum Monte Carlo technique and the stochastic maximum entropy method to study the spectral function of the Hubbard model with the long-ranged Coulomb interaction on the honeycomb lattice.  The emergence of the upper Hubbard band occurs within the semi-metal phase, and we find that the non-local Coulomb interaction enhances the spectral weight transfer to the upper Hubbard band.  We also find that to a first order approximation the addition of the long-range Coulomb interaction does not change the high-energy features.  Upon closer inspection we find that many features can be understood as being governed by Hubbard $U$, while the long-range Coulomb interaction shifts the critical interaction strength $U_c$ to higher values.  Since twisted 2D materials are being envisioned as Hubbard model simulators~\cite{xu_tunable_2022, wu_hubbard_2018}, the results provided here could be used to benchmark the observed properties at charge neutrality.  Finally, we expect that our main conclusion that while the long-range Coulomb interaction does change the critical coupling strength, the formation of local moments that govern the high-energy features remain largely independent of the long-range Coulomb tail and determined mostly by the local Hubbard interaction.  

One could ask how the results presented here shed light on strong correlations in twisted bilayer graphene.  We expect that it should be possible to extend our study of the upper Hubbard band to twisted bilayer graphene to better understand its correlated properties since the non-local interaction is expected to play an important role. Some interesting dynamical properties have already been obtained using the quantum Monte Carlo recently~\cite{Pan2022-ob}, and the method used here could be used to probe the upper Hubbard band.  Moreover, since phonons in this material are expected play an important role~\cite{yudhistira2019gauge}, the influence of the phonon interaction on the upper Hubbard band is also worth investigating.  However, quantum Monte Carlo methods such as those employed in this work have a sign problem away from half-filing.  We emphasize that both in cuprates and in magic angle twisted bilayer graphene, the interesting properties have been observed away from half-filing.  Nonetheless, we believe that the half-filing results can still shed light on the problem.  For example, we believe that our finding that the strongly correlated quantum phase transition is governed by the Hubbard U and not the Coulomb tail, is likely to remain true even away from half-filing.


\section*{Acknowledgement}
This work is supported by Singapore National Research Foundation Investigator Award (NRF-NRFI06-2020-0003), the Singapore Ministry of Education AcRF Tier 2 grant (MOE2019-T2-2-118), Tier 1 grant RG 159/19, and made possible by allocation of computational resources at the Centre for Advanced 2D Materials (CA2DM), and the Singapore National Super Computing Centre (NSCC). HKT thanks the support from the Shenzhen Start-Up Research Funds~(Grant No. HA11409065) and the National Natural Science Foundation of China~(Grant No.  12204130). FFA thanks support from the W\"urzburg-Dresden Cluster of Excellence on Complexity and Topology in Quantum Matter ct.qmat (EXC 2147, project-id 390858490). MU  thanks  the  DFG for financial support  under the  project UL444/2-1.

\bibliographystyle{apsrev4-1.bst}
\bibliography{references}
\end{document}


\title{Supplemental Material for ``Spectral functions of the honeycomb lattice with both the Hubbard and long-range Coulomb Interactions"}

\author{Ho-Kin Tang}
\affiliation{School of Science, Harbin Institute of Technology, Shenzhen, 518055, China}
\affiliation{ Centre for Advanced 2D Materials, National
   University of Singapore, 6 Science Drive 2, Singapore 117546.}
\author{Indra Yudhistira}
 \affiliation{ Centre for Advanced 2D Materials, National
   University of Singapore, 6 Science Drive 2, Singapore 117546.}
 \affiliation{ Department of Physics, Faculty of Science, National University
   of Singapore, 2 Science Drive 3, Singapore 117542.}
   \author{Udvas Chattopadhyay} 
   \affiliation{ Centre for Advanced 2D Materials, National
   University of Singapore, 6 Science Drive 2, Singapore 117546.}
   \affiliation{ Yale-NUS College, 16 College Ave West, Singapore 138527.}
 \author{Maksim Ulybyshev}
 \affiliation{ Institut f\"ur Theoretische Physik und Astrophysik, Universit\"at W\"urzburg, 
   Am Hubland, D-97074 W\"urzburg, Germany.}
 \author{P. Sengupta}
 \affiliation{ Centre for Advanced 2D Materials, National
   University of Singapore, 6 Science Drive 2, Singapore 117546.}
 \affiliation{ School of Physical and Mathematical Sciences, Nanyang Technological 
   University, 21 Nanyang Link, Singapore 637371.}
 \author{F. F. Assaad}
 \affiliation{ Institut f\"ur Theoretische Physik und Astrophysik, Universit\"at W\"urzburg, 
   Am Hubland, D-97074 W\"urzburg, Germany.}
 \author{S. Adam}
 \affiliation{ Centre for Advanced 2D Materials, National
   University of Singapore, 6 Science Drive 2, Singapore 117546.}
 \affiliation{ Department of Physics, Faculty of Science, National University
   of Singapore, 2 Science Drive 3, Singapore 117542.}
 \affiliation{ Yale-NUS College, 16 College Ave West, Singapore 138527.}
 \affiliation{Department of Materials Science and Engineering, 
 National University of Singapore, 9 Engineering Drive 1, 
 Singapore 117575}
\date{\today}
\maketitle

In this Supplemental Material, we present our overall scheme of simulation and analysis. In Sec.~\ref{sec:green}, we discuss the quantum Monte Carlo method and show different scaling of the Green's function. In Sec.~\ref{sec:loren}, we present how we obtain the spectral function using the stochastic maximum entropy method and analyze the spectral function using Lorentzian fitting scheme. In Sec.~\ref{sec:MF}, we present the mean-field calculation shown in Fig.~1 of main text.

\section{Quantum Monte Carlo method and the Green's function at $\Gamma$}
\label{sec:green}
\begin{figure}[b]
    \centering
    \includegraphics[width=1\linewidth]{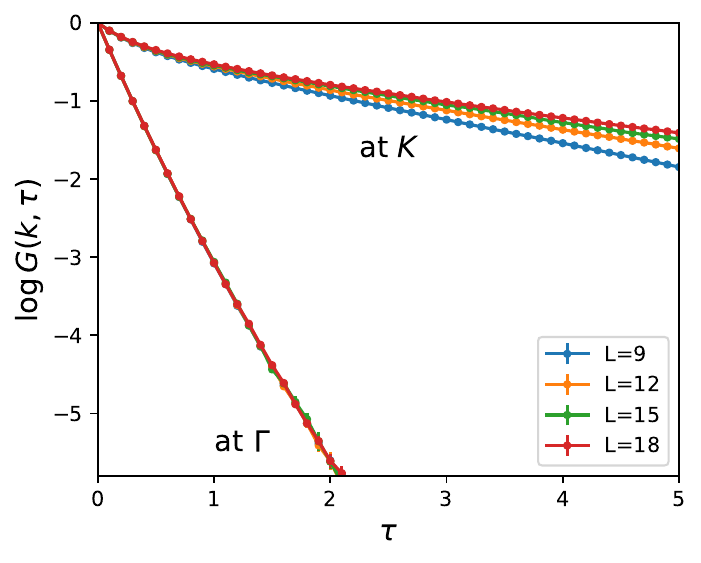}
    \caption{The logarithmic plot of the imaginary time-displaced Green function at the Dirac point $K$ and the $\Gamma$ point of different lattice size $L$, of the pure Hubbard model near the phase transition~($U=4$). The finite size scaling is important at $K$ while the result is mostly independent of the lattice size at $\Gamma$. }
    \label{fig:supp1}
\end{figure}

\begin{figure}[th!]
    \centering
    \includegraphics[width=1\linewidth]{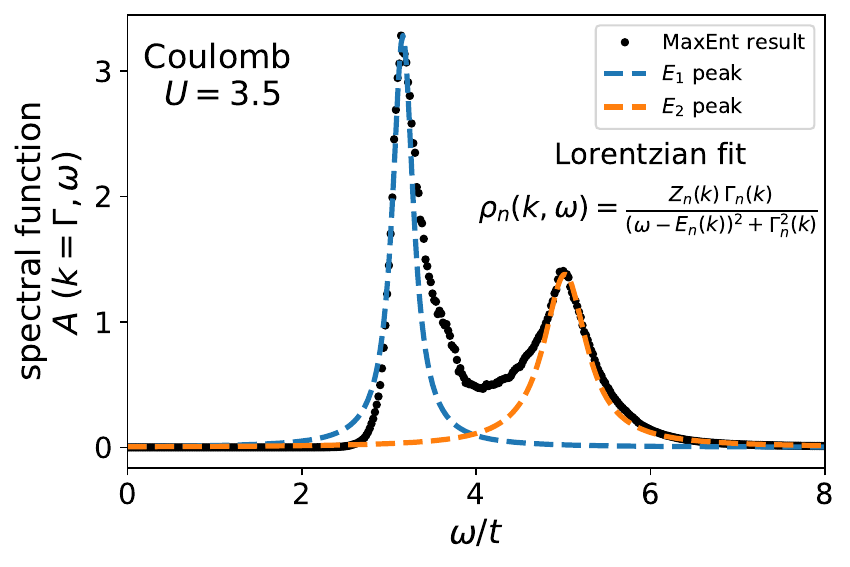}
    \caption{(Black dots) The spectral function $A(k=\Gamma,\omega)$ obtained by the stochastic MaxENT. We fit the two peaks with the Lorentzian function, where they represent the spin polaron peak and the upper Hubbard band respectively.}
    \label{fig:supp2}
\end{figure}

In our study, we use the ground state projective quantum Monte Carlo method~\cite{Assaad2008-qt}, in which we projects out the ground state of the system from a trial wavefunction. 
\begin{equation}
\langle\hat{O}\rangle=\frac{\left\langle\Phi_{0}|\hat{O}| \Phi_{0}\right\rangle}{\left\langle\Phi_{0} \mid \Phi_{0}\right\rangle}=\lim _{\Theta \rightarrow \infty} \frac{\left\langle\Psi_{T}\left|e^{-\Theta \hat{H} / 2} \hat{O} e^{-\Theta \hat{H} / 2}\right| \Psi_{T}\right\rangle}{\left\langle\Psi_{T}\left|e^{-\Theta \hat{H}}\right| \Psi_{T}\right\rangle},\\
\end{equation}

where we use the non-interacting ground state as our trial wavefunction $\Psi_{T}$ and use $\Theta =  40/t$ to project the wave-function into the ground state $\Psi_{0}$. The time interval we used for trotter decomposition is $0.1/t$. To include the non-local Coulomb interaction, we use the continuous type of the Hubbard Stratonovich transformation~\cite{} to decouple the fermionic interaction term into the auxiliary field coupled to the fermion. After the Monte Carlo sampling, we can obtain the imaginary time displaced Green's function,
$G(k,\tau) \equiv\left\langle\Phi_{0}\left|\hat{c}_{k}^{\dagger}(\tau) \hat{c}_{k}(0)\right| \Phi_{0}\right\rangle$. Near the phase transition, the finite size effect is severe at the Dirac point $K$, while the sacling is much more stable at $\Gamma$ point. The logarithmic plot of the Green's function obtained near the phase transition is shown in Fig.~\ref{fig:supp2}. Our analysis is mainly based on $L=15$ result at the $\Gamma$ point in the main text.

\section{Stochastic MaxEnt method and the Lorentzian fitting}
\label{sec:loren}

The spectral function $A(k, \omega)$ is obtained by doing the analytical continuation of the following equation.
\begin{align}
\label{eq:inv}
    G(k,\tau)&= \frac{1}{\pi} \int d\omega K(\tau,\omega) A(k,\omega),
\end{align}
where $K(\tau,\omega)= e^{-\tau \omega}$ is the transformation kernel at zero temperature. Obtaining $A(k,\omega)$ from the inversion is an ill-posed problem, as many possible solution is available for an given Green's function. One of the most reliable ways is to use the stochastic maximum entropy method~\cite{smem}. The method obtained great success in dealing with many types of the spectral function corresponding to the given Green's function. The code is available in a open-source library of the algorithms for lattice fermions~(ALF)~\cite{ALF1,ALF2}.

Once we obtain the spectral function at the $\Gamma$ point, we fit the Lorentzian function $\rho_n(k, \omega)=\frac{Z_n(k)\,\Gamma_n(k)}{\left(\omega-E_{n}(k)\right)^{2}+\Gamma_n^{2}(k)}$
 to the observed peaks, where $n$ is the band index. In the weakly interacting regime, only one quasiparticle peak is observed, so we fit one Lorentzian function. Near the phase transition, two distinguishable peaks appear in $A(\Gamma,\omega)$, so we use two Lorentzian function to perform the analysis. An example is given in Fig.~\ref{fig:supp2}. We obtain the excitation energy $E_n$, the weight $Z(k)$, and the lifetime $\tau(k)=1/(2\Gamma(k))$ from the analysis.

\section{Mean-field calculation of Spectral Function}
\label{sec:MF}
The hamiltonian in presence of on-site and nearest neighbor Coulomb interaction reads
\begin{equation}
    \hat{H} = -t \sum_{\langle i,j \rangle, \sigma } 
    (\hat{a}^\dagger_{i,\sigma} \hat{b}_{j, \sigma} + \textrm{h.c.}) +
    U\sum_{i} n_{i, \uparrow} n_{i, \downarrow} +
    W \sum_{\langle ij \rangle} \hat{n}_i \hat{n}_j
\end{equation}

\noindent where $\langle \rangle$ denotes nearest neighbor sites and $\hat{n}_i = \hat{n}_{i, \uparrow} + \hat{n}_{i, \downarrow}$. The interaction terms are solved using the mean field approximation $n_{i, \sigma} = \langle n_{i, \sigma} \rangle + \delta n_{i, \sigma}$, where $\langle n_{i\sigma} \rangle$ is average density at site $i$ spin $\sigma$ and $\delta n_{i\sigma}$ is deviation from mean values. Only linear orders in fluctuations $\delta n$ are retained:

\begin{align}\label{HubUMF}
    H_U \approx \frac{U}{2} \sum_{i, \sigma} \left[ 
    \langle n_{i,\sigma} \rangle \langle n_{i, \bar{\sigma}} \rangle  + \delta n_{i, \sigma} \langle n_{i, \bar{\sigma}} \rangle + \delta n_{i,\bar{\sigma} } \langle n_{i,\sigma} \rangle \right] \\
    =U\sum_{i,\sigma}n_{i,\sigma} \langle n_{i,\bar{\sigma}} \rangle - U\sum_i \langle n_{i,\uparrow} \rangle \langle n_{i,\downarrow} \rangle 
\end{align}
Similarly for the nearest-neighbor Coulomb interaction
\begin{equation}\label{NNMF}
    H_W \approx W \left[ \sum_{\langle ij \rangle, \sigma } n_{i\sigma} \langle n_j \rangle + n_{j\sigma} \langle n_i \rangle - \langle n_i \rangle \langle n_j \rangle  \right]
\end{equation}
where $\langle n_i \rangle = \langle n_{i\uparrow} \rangle + \langle n_{i\downarrow} \rangle$.
The mean -field hamiltonian is then solved with the densities calculated self-consistently for a given filling and the retarded Green function is calculated:
\begin{equation}
    G_\sigma (\Vec{k}, \omega) = \frac{1}{\omega + i0^+ - H_{\sigma}(\Vec{k})}
\end{equation}
and spectral function is calculated as
\begin{equation}
    A(\Vec{k}, \omega) = -\textrm{Im} \sum_{\sigma} \textrm{Tr} G_\sigma (\Vec{k}, \omega) 
\end{equation}.

\noindent In the Fig.~1, the black solid line represents the spectral function calculated using mean field theory. The $U$ values used are different from QMC data to account for the fact that the critical value $U_c$ of transition to AFM phase is smaller for MF result: $U_c^{MF}/t \approx 2.23; U_c^{QMC}/t \approx 3.78 $.